\documentstyle[12pt]{article}
\begin{document}

\begin{center}
{\Large Celestial Origin of the Extended EUV Emission \\
from the Abell 2199 and Abell 1795 Clusters of Galaxies}
\vspace{5mm}

{Richard Lieu$\,^{1}$ , Jonathan P. D. Mittaz$\,^{2}$, 
Massimiliano Bonamente$\,^{1}$, \\ 
Florence Durret$\,^{3}$, Sergio Dos Santos $\,^{3}$ and Jelle S. Kaastra$\,^{4}$}

\vspace{5mm}

{\footnotesize \(^{\scriptstyle 1} \){Department of Physics, University of Alabama,
Huntsville, AL 35899, U.S.A.}\\
\(^{\scriptstyle 2} \){Mullard Space Science Laboratory, UCL,
Holmbury St. Mary, \\ 
Dorking, Surrey, RH5 6NT, U.K.}\\
\(^{\scriptstyle 3} \){Institut d'Astrophysique de Paris, 
CNRS, 98bis Bd Arago, F-75014 Paris, France}\\
\(^{\scriptstyle 4} \){SRON Laboratory for Space Research, Sorbonnelaan 2,\\
 NL-3584 CA Utrecht, The Netherlands}\\
}

\end{center}

\abstract{Several authors (S. Bowyer, T. Bergh\"offer, and E. Korpela,
hereinafter abbreviated as BBK)
recently announced that the luminous extended
EUV radiation from the clusters Abell 1795 and Abell 2199, which
represents the large scale presence of a new and very soft 
emission component, is an illusion in the
EUVE Deep Survey (DS) detector image.
Specifically BBK found that the radial profile of
photon background surface brightness,
for concentric annuli centered at a detector position
which has been used to observe cluster targets, 
shows an intrinsic
`hump' of excess at small radii which resembles the detection of
extended cluster EUV.  We accordingly profiled background data, but found
no evidence of significant central excess.  However, to avoid argument
concerning possible variability in the background pattern, we performed
a clincher test which demonstrates that a cluster's EUV profile is
{\it invariant with respect to photon background}.  The test involves
re-observation of A2199  and A1795
when the photon background was respectively three
and two times higher than before, and
using a different part of the detector.  The radial profiles of both
clusters, which have entirely different shapes, were
accurately reproduced.  In this way the BBK scenario is 
quantitatively excluded, with the inevitable conclusion that
the detected signals are genuinely
celestial.}

\vspace{5mm}
\noindent
{\bf 1. Introduction}

\vspace{5mm}
Since the original EUVE discovery (Lieu et al 1996a) of the CSE effect,
substantial
development in the field has taken place, including the
ROSAT and BeppoSAX confirmatory detection of
soft X-ray (0.1 - 0.4 keV) excesses (Lieu et al 1996a,b;
Bowyer, Lampton and Lieu 1996; Fabian 1996; Mittaz, Lieu and Lockman 1998;
Bowyer, Lieu and Mittaz 1998; Kaastra 1998, Kaastra et al 1999)
from Virgo, Coma,
A1795, A2199, and A4038, and the proposition of theoretical ideas
which relate the CSE
to other new
radiation components (in particular the hard excess (HEX) emission
recently found from some clusters;  details and
references may be found in Ensslin and Biermann 1998, Sarazin and Lieu 1998, 
Rephaeli et el 1999, Kaastra 1998 and Fusco-Femiano et al 1998).

Despite the exciting progress, some
researchers continue to question the reality of
CSE. 
In a recent workshop on clusters of galaxies 
(Ringberg 1999\footnote{Thermal and relativistic
plasmas in clusters of galaxies, Schl\"oss Ringberg, Germany,
April 19-23, 1999.}) BBK suggested
that the detection of spatially
extended EUV emission from the clusters A1795 and A2199
(Mittaz, Lieu, and Lockman 1998, Lieu, Bonamente, and Mittaz 1999), which
implies a radially increasing importance of the CSE,
results from an error in the photon background
subtraction procedure.  

The original authors of the CSE
detection determined the
background level from a large outer annulus 
(typically between  15 abd 30 arcmin cluster 
angular radii) of the EUVE Deep Survey
(DS) Lex/B (69 - 190 eV) image, a region where the radial profile of
the cluster surface brightness has reached asymptotic flatness (i.e. the
brightness no longer falls with radius).  BBK, on the other hand,
asserted that the 
photon background radial profile is enhanced from the asymptotic
level within a circle of radius $\sim$ 15 arcmin centered at a position
of the detector where observations of clusters were performed
(the enhancement is in the form of a `hump' which peaks at
$\sim$ 5 arcmin radius);
outside this circle
the profile does not flatten, but gradually decreases with radius.
Thus the original usage of an asymptotic background would have led to
an undersubtraction effect in the central 15 arcmin region, and hence
a false detection of cluster EUV.

BBK then applied the
forementioned photon background profile as a `template' (i.e. the
shape remains unchanged and the `normalization' is adjusted to obtain
agreement  with the measured photon
background of a cluster field at large radii)
in a `revised' background subtraction procedure which led to the removal
of all previously detected cluster signals from A1795 and A2199 at radii
beyond $\sim$ 5 arcmin.  Moreover within 5 arcmin the CSE was turned
into an intrinsic cluster absorption effect, with the detected brightness
being {\it smaller} than the amount of EUV expected from the hot intracluster
medium (ICM).

\begin{figure*}[h]
\vspace{0cm}
\includegraphics{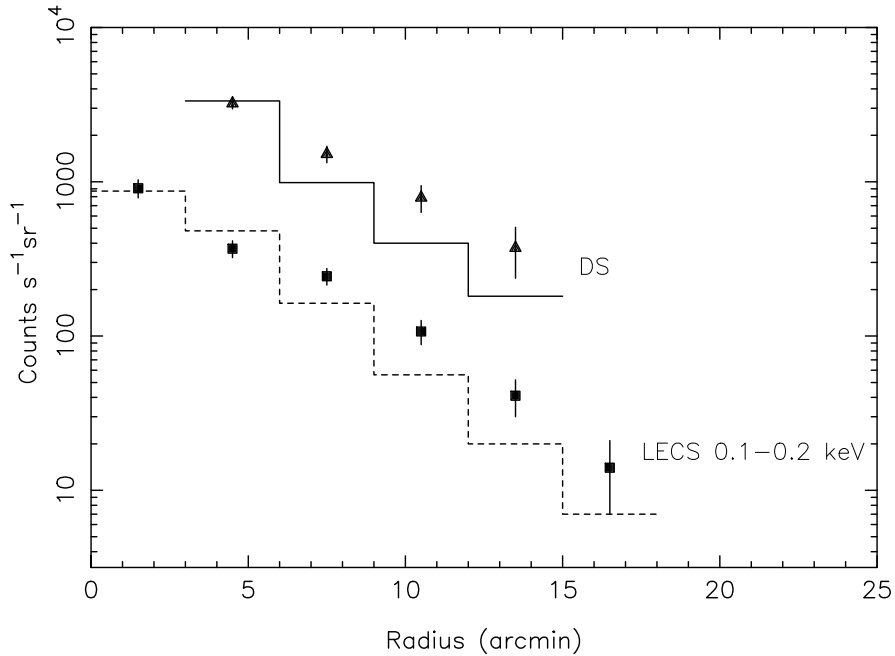}
%\special{psfile=fig_sax.ps  hoffset=200 voffset=-10 hscale=35 
%vscale=35 angle=-90}
\vspace{8.5cm}
\caption{\footnotesize A2199 surface brightness profile in the 0.1 - 0.2 keV band as
observed by the EUVE/DS and SAX/LECS detectors; the solid and dashed
lines give the respective expected DS and LECS brightness had the soft
emission been due only to the hot ICM gas radiation, thereby revealing
CSE at the outer portions of the cluster.}
\end{figure*}

\newpage
\noindent
{\bf 2. No CSE for A1795 and A2199 ?}
\vspace{5mm}

The results of BBK raise several perplexing questions.

\noindent
(a) The CSE effect was confirmed by the LECS instrument aboard BeppoSAX
(Kaastra et al 1999, see also Figure 1).  The radially rising trend
of the CSE, as detected by LECS, is commensurate with that found
by EUVE/DS (Figure 1) when instrument effective areas are taken into account.
Moreover the LECS soft excess was noted to exist only below
0.2 keV, again in agreement with the conclusion obtained
from simultaneous modeling of the EUVE/DS and ROSAT/PSPC data of this
cluster.

\noindent
(b) A multi-scale wavelet analysis
of the EUVE/DS data of A1795 (Durret et al 1999) shows
clear signatures of cluster emission out to a radius of at least 8 arcmin
with the isophot levels rising radially to
reach a factor of $\sim$ 4 above
those expected from the hot ICM (see Figure 2).  The same technique as
applied to a background field of comparable exposure 
(see next section for details of this field) revealed no
statistically significant features at 
$\geq$ 3 $\sigma$ level around detector areas
where clusters were normally observed by the DS.  To facilitate
direct appreciation by the reader of these points, FITS images of 
A1795 (90 ksec exposure), A2199 (48 ksec) and background (85 ksec)
can be downloaded via anonymous ftp to ftp://cspar.uah.edu/input/max
(the README.TXT file contains further information).  These images
are raw data (for precise meaning see section 3) which underwent
only one stage of additional processing: they were smoothed with
a constant gaussian filter (size commensurate with the DS point spread
function) to enhance larger features.   Each image has equal spatial scale,
and the central portion corresponds to the same part of the DS detector.
It will immediately be
obvious to the viewer that the cluster fields 
exhibit a luminous
extended glow (with A2199 particularly sprawling) which is {\it not}
reproduced in the background field.

\begin{figure*}[h]
\vspace{1cm}
\includegraphics{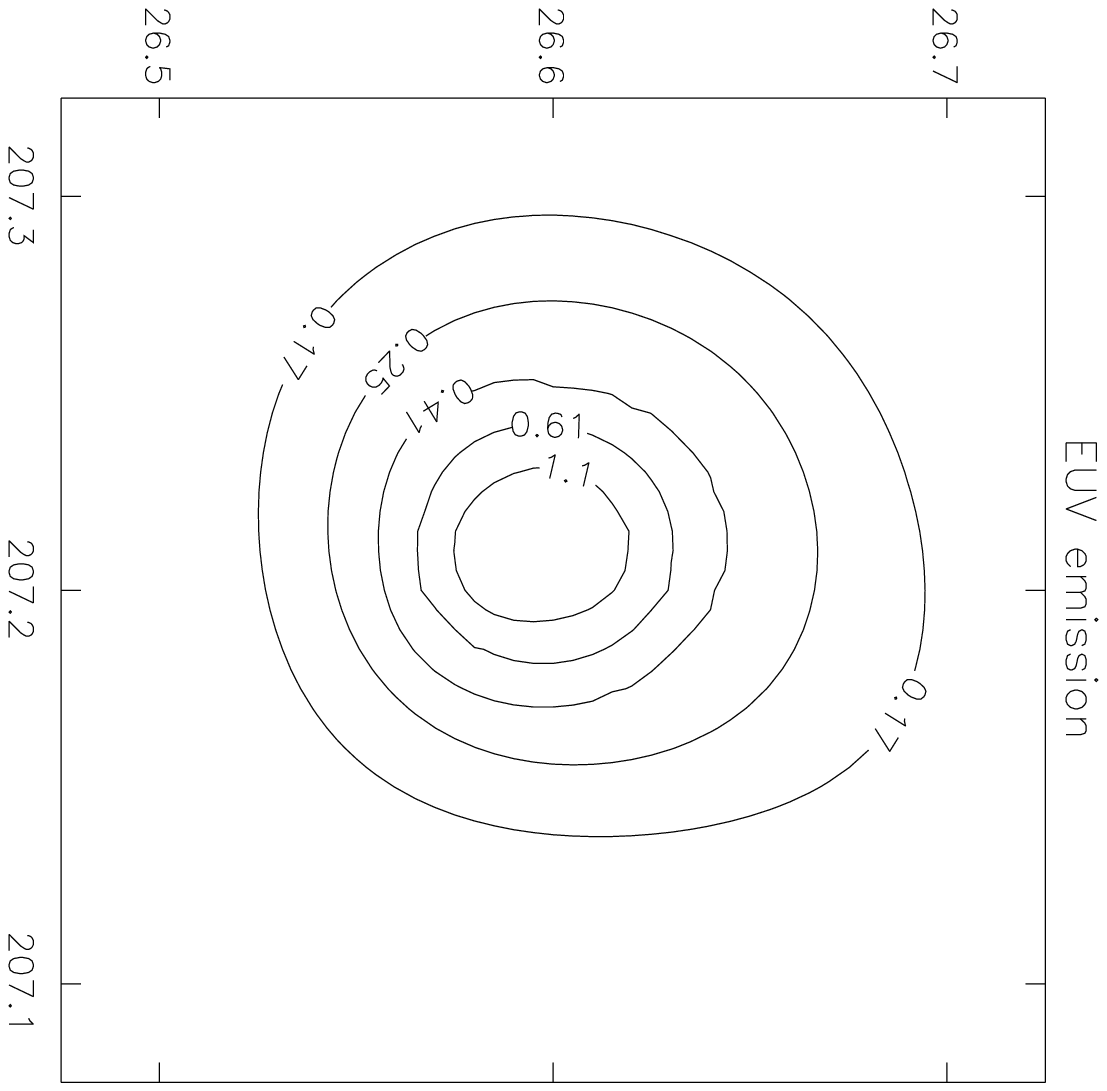}
\includegraphics{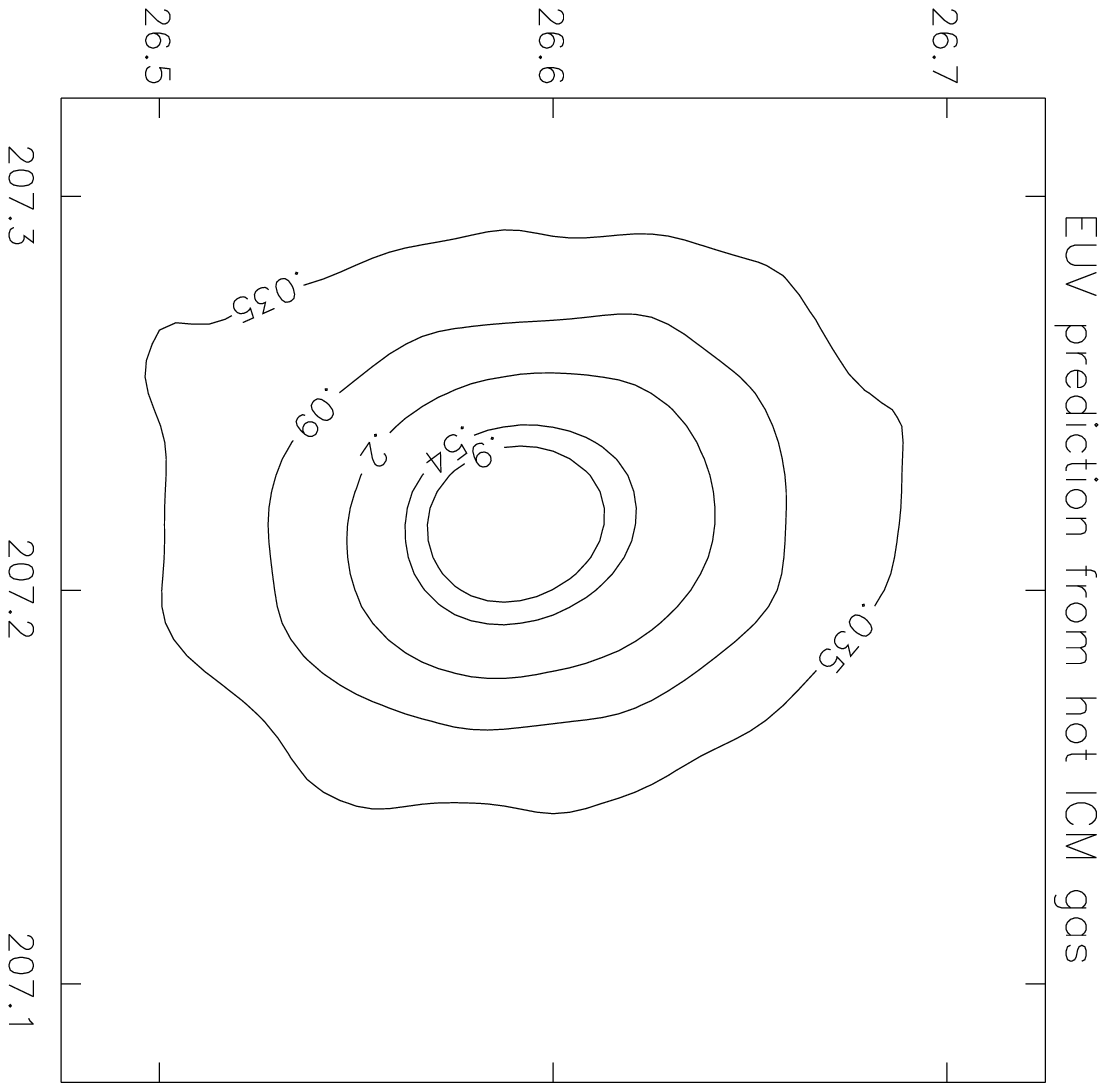}
\vspace{6.0cm}
\caption{\footnotesize Left: EUV surface brightness contours of A1795 in units of
photons ksec$^{-1}$ arcmin$^{-2}$, as obtained by a multi-scale wavelet
analysis.  All isophots are $\geq$ 3 $ \sigma $ significant.
Right: Predicted EUV emission from the hot ICM gas; this
map was obtained by fitting single temperature plasma models to
the ROSAT/PSPC spectral data for different regions of the X-ray
image.  It is evident that the observed values are far in excess of
the predicted ones, the excess becoming stronger with
increasing cluster
radius. Further, the Poisson noise of the EUVE/DS background is
such that if all the EUV were from the hot ICM, no detection would
be possible beyond 5 arcmin radius, i.e. any signal beyond this
radius immediately implies CSE.}
\end{figure*}

\noindent
(c) The additional background subtracted by BBK did not lead to the
removal of CSE from the Virgo and Coma cluster data.  A natural puzzle
is why these two clusters exhibit CSE but not the others (in fact the rest
suffer from the opposite effect: they are strongly intrinsically absorbed).
We note in this regard
that the presence of CSE in Virgo and Coma excludes the possibility of a
simple correlation or anti-correlation between the CSE on one hand, and
cooling flow or merging/subclustering on the other.

\begin{figure*}[h]
\vspace{2cm}
\includegraphics{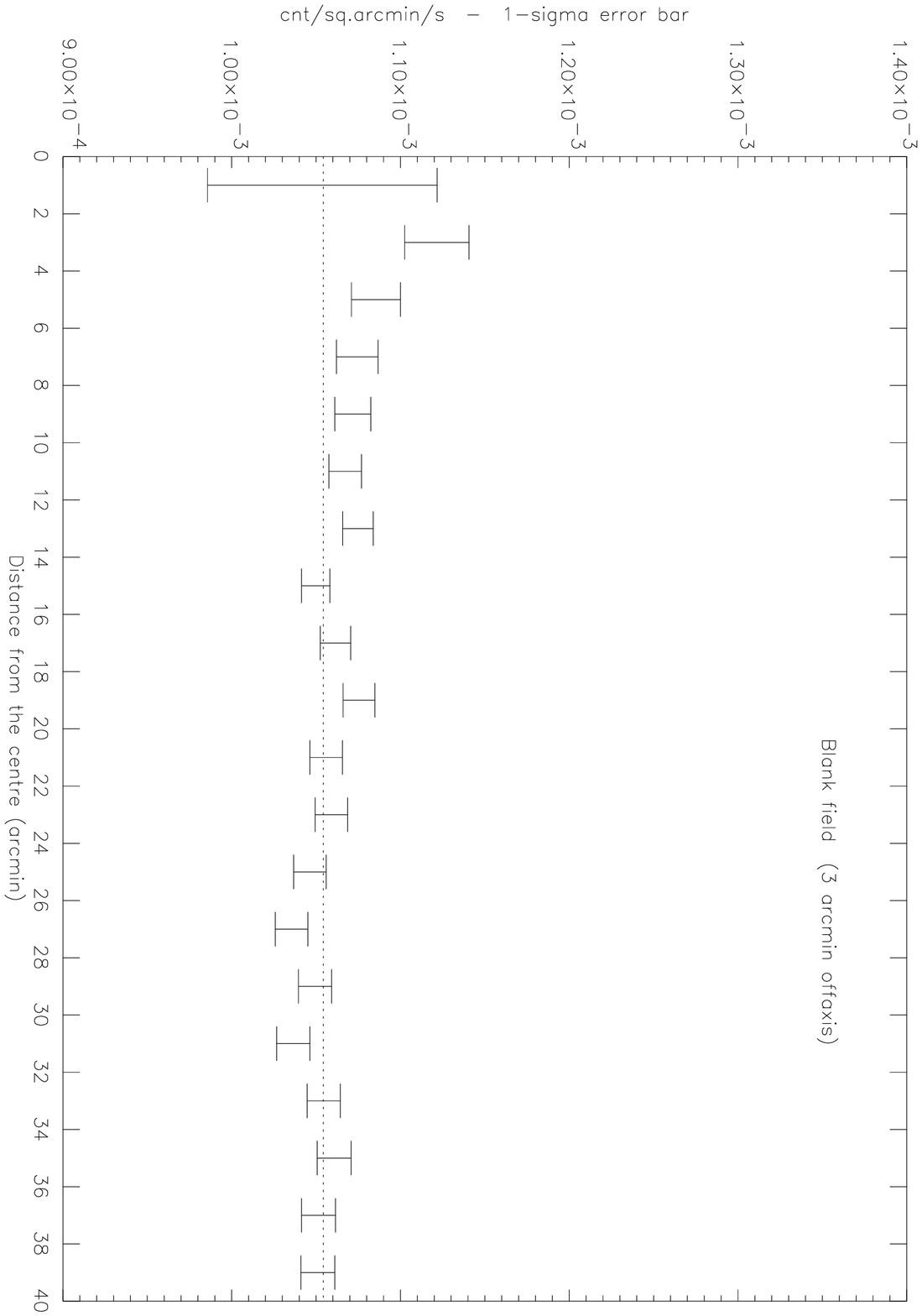}
\includegraphics{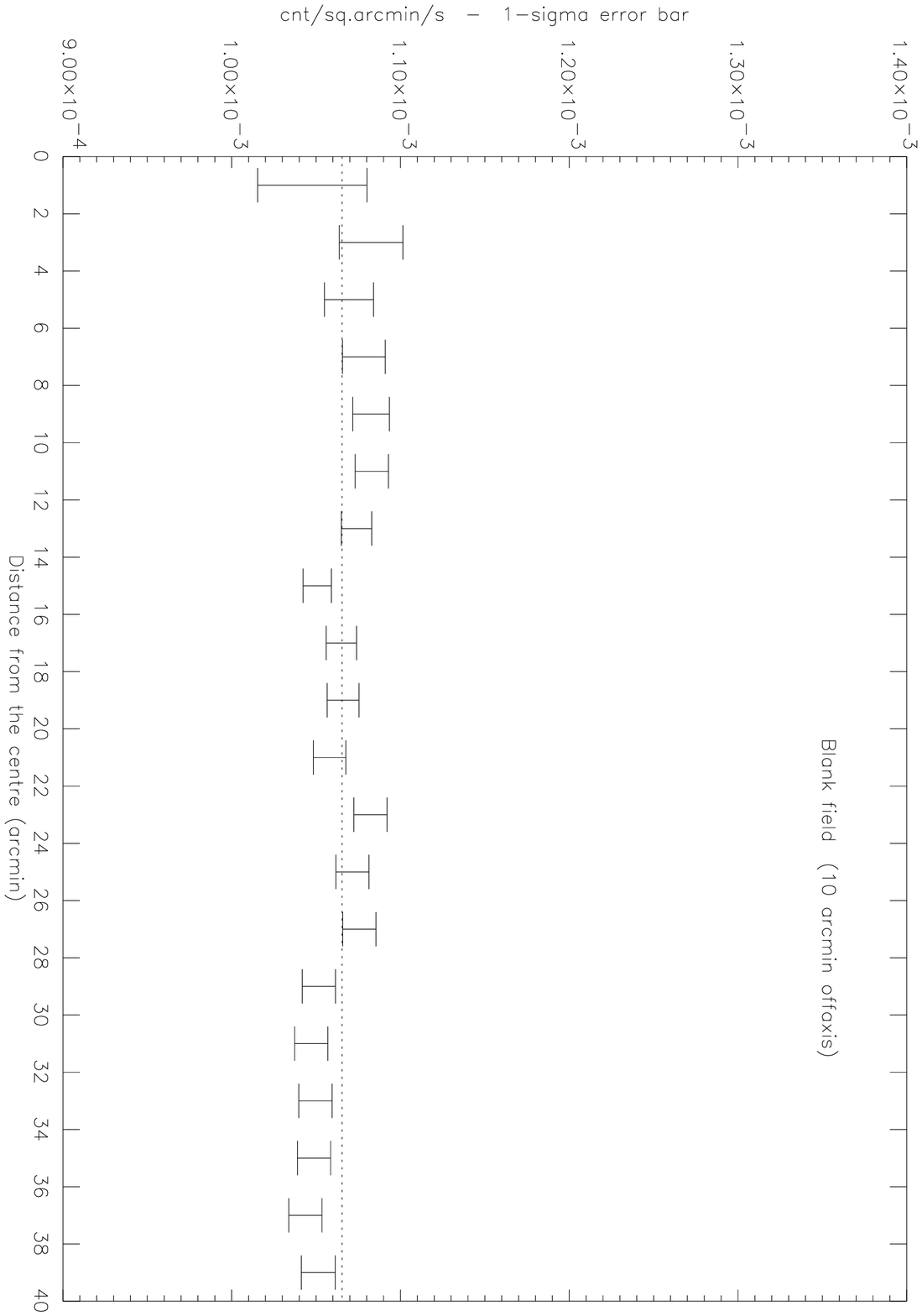}
\vspace{5cm}
\caption{\footnotesize Left: Radial profile of EUVE/DS background with center at
$\sim$ 3 arcmin off-axis.  Right: same as left panel, except center is
now at $\sim$ 10 arcmin off-axis.  In each case the dotted line
represents the average brightness between 15 and 30 arcmin (a region
typically chosen to determine the background of cluster fields).}
\end{figure*}

\noindent
(d) The strong intrinsic absorption of A1795 and A2199 within cluster
radii of $\sim$ 5 arcmin should have measurable effects in the ROSAT/PSPC
0.25 keV band and in the LECS data at energies $\geq$ 0.2 keV, yet it is not
immediately obvious why
such effects are totally absent from the PSPC and LECS data of A1795
and A2199.

At the request of workshop
participants, including some of the scientific
organizers, we are circulating this memo to address the BBK criticism
of the CSE.
In particular we shall demonstrate that the
EUVE/DS photon background does not exhibit a template radial distribution
involving excess counts at small radii.  In fact the integrity of
the original method of asymptotic background 
subtraction is assured by pairs of observations 
of the same clusters, under conditions of very different photon
background levels,
yielding the {\it same} radial brightness dependence
which varies only from cluster to cluster.

\vspace{5mm}
\noindent
{\bf 3. Background radial profiles from blank fields}
\vspace{5mm}

Public domain EUVE DS data are accessible from HEASARC, and standard
data products (hereinafter referred to as raw data)
are obtained by running routine pipeline software packages which
accept satellite telemetry as input.  The raw data (with a
background that includes photons, particles, and detector
intrinsic noise) can further
be processed in any number of ways, some of which could lead to
artifacts.  The CSE results published so far are,
however, based on the
analysis of raw data.
We show in Figure 3 the radial profiles
of the DS  background, obtained with the center at $\sim \;$3 and 
$\sim$ 10 arcmin
off-axis\footnote{the former is the position where clusters centroids were
located during the cycle 4 and 5 observations, while the latter is 
the position for the cycle 6
re-observations of, e.g., A2199.} along the detector x-axis, i.e. the
direction parallel to the long side of the relevant (Lex/B) filter.
The data were gathered by merging three blank
field\footnote{Defined as fields which do not contain bright sources.
The three specific datasets correspond to the targets 2EUVE J1100+34.4,
2EUVE J0908+32.6, and M15.}
pointings which took place in July 1994 and December 1996, with a total
exposure of $\sim$ 85 ksec, comparable to the longest cluster observation
by EUVE.  It can be seen from Figure 3 that the profiles
reveal no significant central enhancement - there are no signatures of
extended emission resembling those of A1795 (Figure 1 of Mittaz et al 1998)
or A2199 (Figure 1 of Lieu, Bonamente and Mittaz 1999).  Specifically
for the left plot the average 0 - 15 arcmin background is
1.75 $\pm$ 0.50 \% higher than the 15 - 30 arcmin value, and for
the right plot the same comparison yields 0.9 $\pm$ 0.5 \%.

\begin{figure*}[h]
\vspace{1cm}
\includegraphics{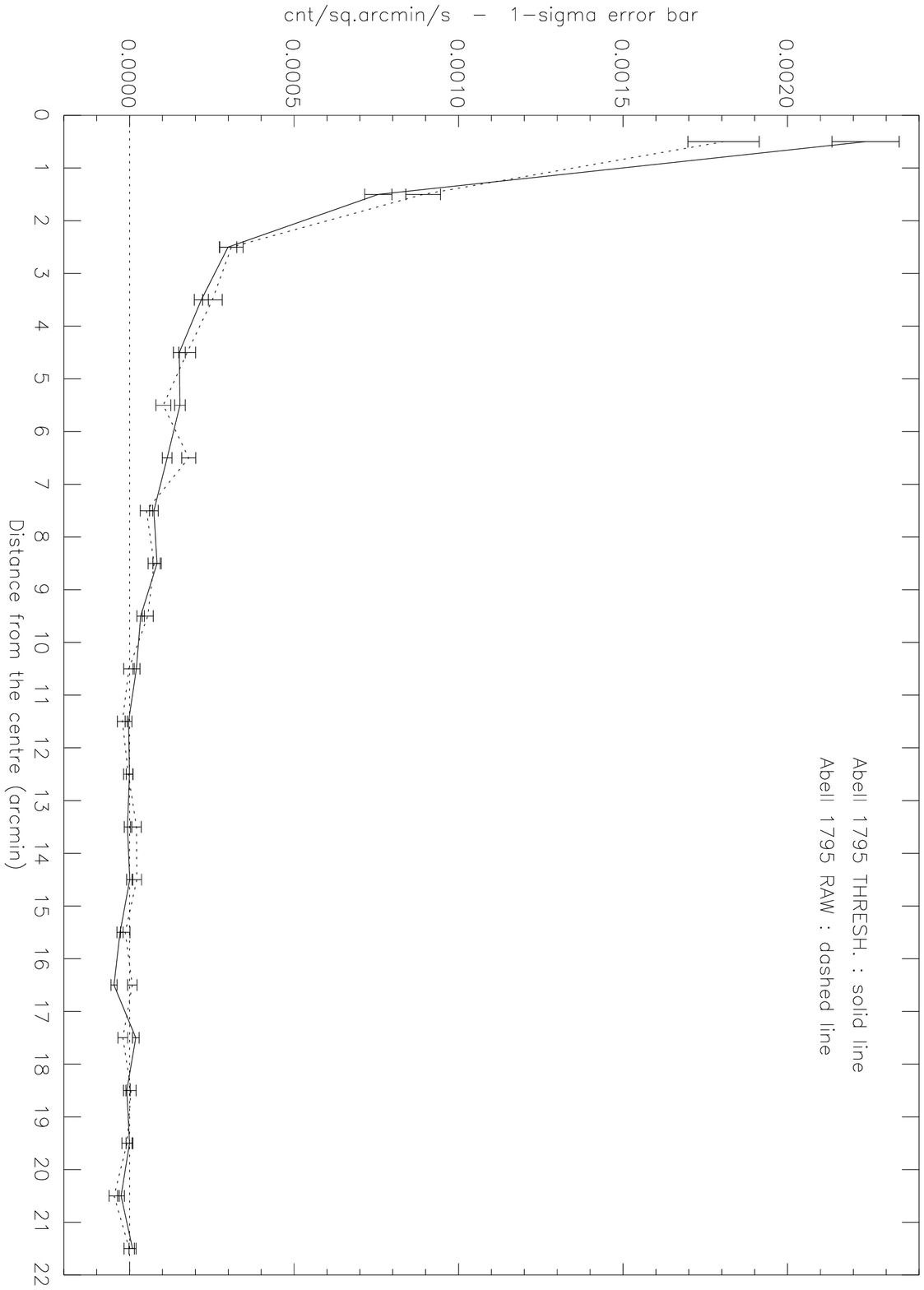}
\includegraphics{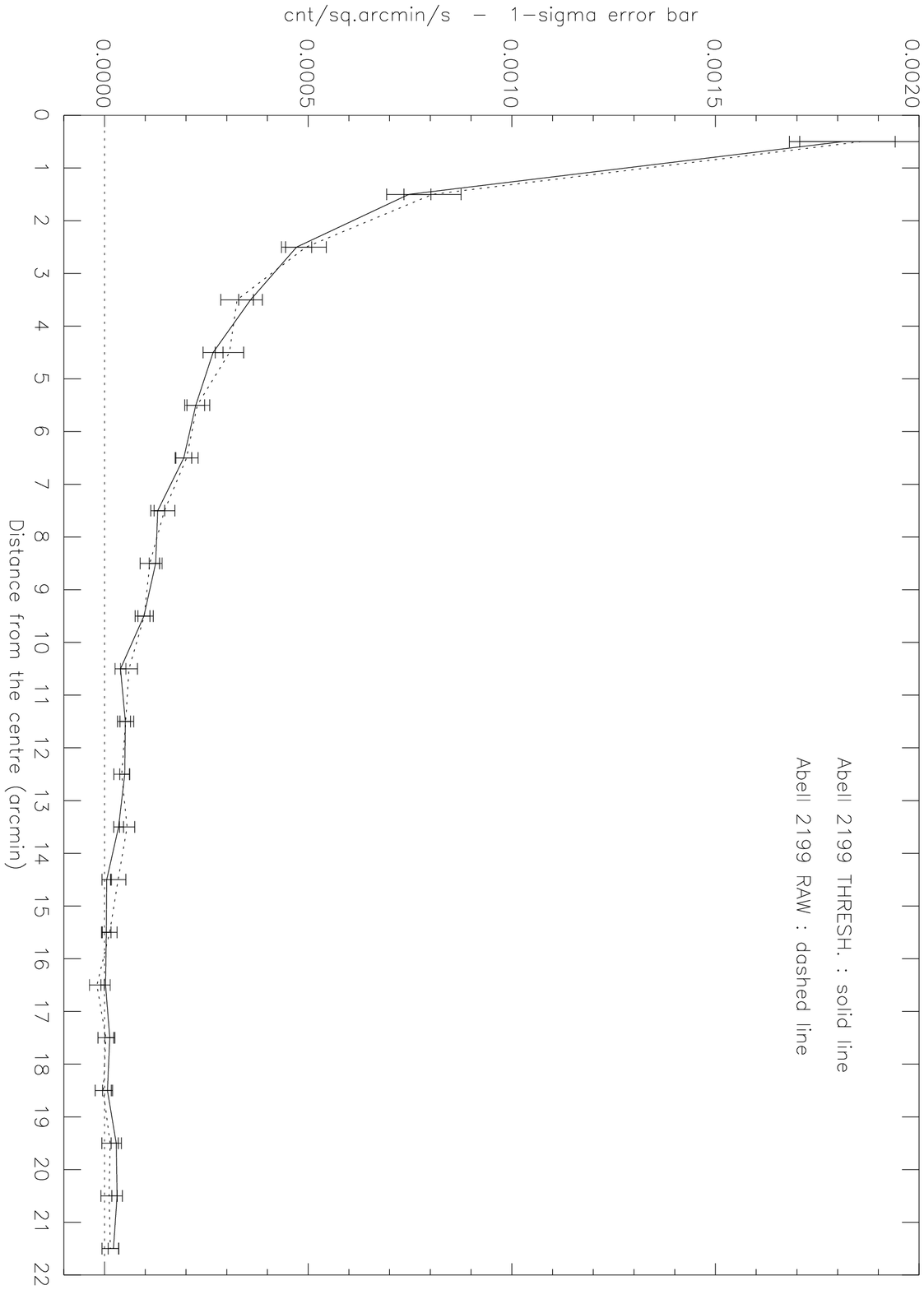}
\vspace{6.5cm}
\caption{\footnotesize Left: Background subtracted
radial profiles of raw and photon (PH thresholded) data from an
EUVE/DS observation of A1795.  Right: same as left, now for A2199. Point sources and 
detector artifact contributions were removed.
Note in each case the excellent agreement between the profiles.}
\end{figure*}

A technique often applied to reduce the raw background 
by removing its particle component in the
cluster signal is {\it pulse height (PH) thresholding}.  It has
the positive effect of lowering the Poisson noise in the
cluster signal.
In brief, one first
constructs a PH
histogram of all detected counts, which
consists of a normally distributed peak of photon events superposed
upon an underlying (pseudo power-law) continuum of particle events.
By suitably setting the PH thresholds to select only events within the
photon peak, it is possible to reject the majority of the particle
background without compromising the cluster signal.  The precise
fractional reduction of the particle component varies from one
dataset to the next.  As an example, for the first observation of
A2199 at least $\sim$ 62 \% of the background was due to particles,
and was removed by PH thresholding.
The resulting product, from which BBK derived the photon
background,
is no longer a raw dataset.  However, if proper processing is
performed the
integrity of the PH thresholded data may still be maintained, as
demonstrated by Figure 4 where there is close
resemblance between the radial profiles of the raw and
PH thresholded data (with no extra features in the latter)
for the clusters A1795 and A2199.
We also note in passing that, like the raw profile
and unlike the BBK template, the 
thresholded photon background is
flat - it does not exhibit BBK's downward slope from
10 arcmin outwards (Fig. 4, left).  Thus such a pattern, even if it were to exist,
is not the norm of behavior, and certainly does not have to be
be present in every observation.

\begin{figure*}[h]
\vspace{1cm}
%\special{psfile=a1795_bowyer_lieu.ps  hoffset=420 voffset=-200 hscale=50 vscale=50 angle=90}
\includegraphics{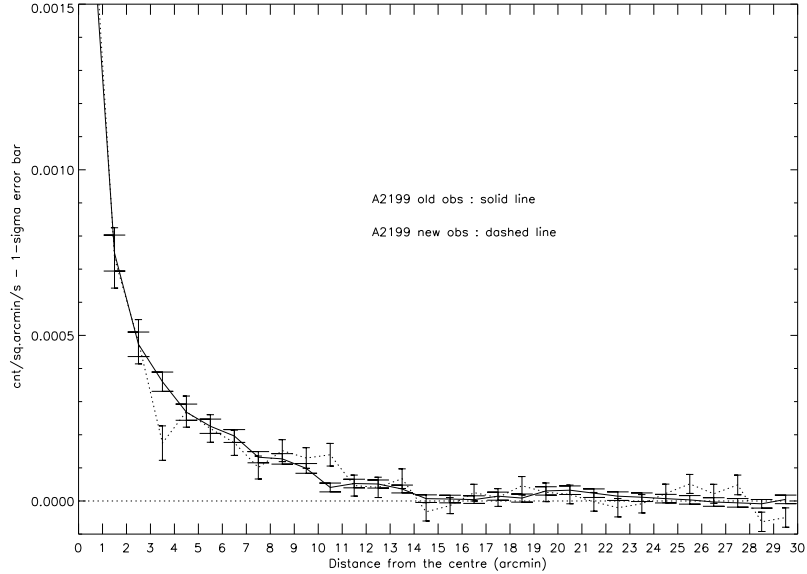}
\vspace{7cm}
\caption{\footnotesize PH thresholded and background subtracted A2199 radial profiles
as obtained by the original observation (Fig. 4) and another
observation which took place approximately one year later; all point sources and detector artifact contributions were removed.  Note again
the consistency of the two profiles.}
\end{figure*}

\vspace{5mm}
\noindent
{\bf 4. Still, does this mean Lieu et al were right ?}
\vspace{5mm}

At the very least, therefore,
it is clear that the DS background 
profile is not a template which carries features reminiscent of
a cluster detection, as advocated by BBK.  On the other hand,
BBK did prompt an
intruiging question, viz. given
that the EUVE/DS background (like that of most satellite detectors) is
a complex function of many parameters,
how can one {\it guarantee} that the original
asymptotic subtraction procedure
is correct, short of actually measuring the background underlying
each cluster at the time of every observation (an impossible
task) ?

Fortunately there is a clear answer to even this question.  The test
is essentially an extension of the approach depicted in Figure 4.

\begin{figure*}[h]
\vspace{1cm}
%\special{psfile=1795_percent.ps  hoffset=420 voffset=-240 hscale=45 vscale=45 angle=90}
\includegraphics{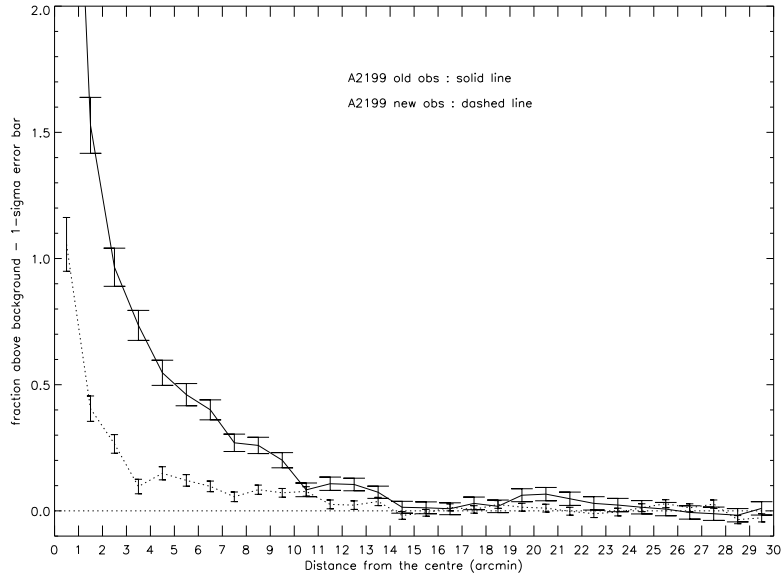}
\vspace{7cm}
\caption{\footnotesize Radial profile of the two observations of A2199 with the surface
brightness expressed as fraction above the asymptotic 
photon (i.e. PH thresholded) background. Again point sources and detector artifact contributions were removed. Note
that, unlike the previous figure, there is now a
large discrepancy between the profiles within a radius of 13 arcmin,
meaning that cluster extended EUV
cannot be modeled as a template variation of the photon background.}
\end{figure*}

\begin{figure*}[h]
\vspace{1cm}
\includegraphics{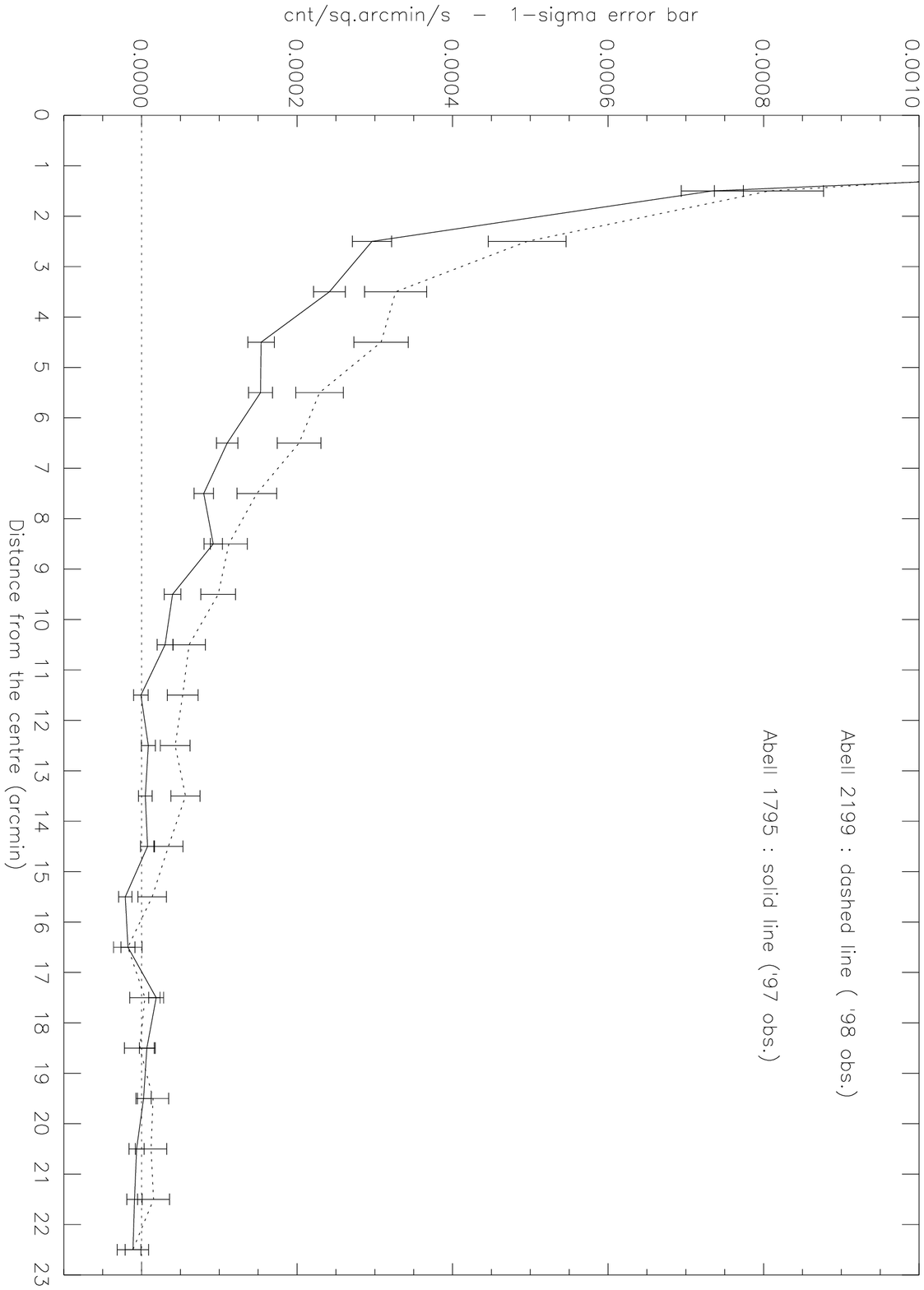}
\vspace{7cm}
\caption{\footnotesize Radial profile of the EUV surface brightness of A2199
and A1795 indicating very different spatial distribution of radiation 
( all point sources and detector artifacts were removed) .
Note that the detection limit radius is $\sim$ 10 arcmin for A1795, but
$\sim$ 15 arcmin for A2199.}
\end{figure*}

\vspace{5mm}
\noindent
{\bf 5. Clincher test of background subtraction -
the reproducibility of cluster EUV signals}

\vspace{5mm}
The longevity of EUVE made it possible for repeated observations
of the same clusters - indeed some of these took place within cycle 6.
The DS background of the re-observations is in general considerably
higher (typically by a factor of $\sim$ 2)
than that of the original data, {\it mainly due to an increase in the
photon background}.
Further, to explore the effects of detector
uniformity the clusters were usually re-observed in a different detector
position, some 10 arcmin off-axis as described earlier.  Thus, e.g., in
the case of A2199 a comparison using
PH thresholded data
reveals that during re-observation the photon background 
was $\sim$ three times higher than before.

New data for
three bright clusters: A2199, A1795, and Coma 
are available, and in {\it every} case the signals
peter out to a flat background with a radial profile
which is consistent with that of the original observation.
As an example, we show in Figure 5 the PH thresholded and
background subtracted profiles
of A2199.

The significance of this agreement,
however, lies in the fact that the large difference in the
photon background level between each pair of observations
{\it clinches}
any scenario which attributes cluster signals to a  
centrally enhanced template profile of the
photon background.
To prove our statement, Figure 6 shows the radial profiles in
the form of
percentage above this background\footnote{Data
for Coma and A1795 only reinforce
the same points
made by Figures 5 \& 6.  They are therefore not
shown here, except to add that for A1795 BBK also noted the 
reproducibility of the radial profile, and that the re-observation
photon background is twice as high as before, again quantitatively
excluding any part of the cluster detection as a background effect}.

It is clear that within the context of
the BBK scenario
the photon background must
assume {\it two} templates, suitably correlated with each other as
to produce the same absolute brightness profile (Figure 5) !
In particular, if the 5 - 15 arcmin signals were to be a background
variation effect (as advocated by BBK) this variation will have to form
templates (Figure 6) which are statistically
distinct, and which conspired\footnote{The conspiracy could exist in a `relatively
simple' form if the higher re-observational background is due 
primarily to an
increase in the level of a `flat profile' component, such as particles,
superposed upon a photon component which is
constant, and which carries an intrinsic
enhancement at 
the inner radii where cluster EUV were reported.  However, this
scenario is not viable because Figures 5 and 6 refer to thresholded
background, which consists principally of photons (see earlier discussion).
Certainly particles cannot be responsible for
the 3-fold increase of the background in the new A2199
data.} to yield Figure 5.

\vspace{5mm}
\noindent
{\bf 6. Another cosmic conspiracy}

\vspace{5mm}
Finally, suppose we {\it do} accept the last two statements of
the previous section as truth,
do we need even more cosmic
coincidences to explain any `loose ends' that may still exist ?
If yes, then perhaps one can safely declare that 
the premise of CSE as
a background variation effect has been given sufficient `benefit
of the doubt', and exclude it as a sensible, viable approach ?

The answer is indeed yes, for if there exists
highly contrived `multiple templates' which correspond to
a single absolute brightness profile irrespective of the background level
(the `beyond BBK' scenario described in section 4), one will
be forced to conclude that such a profile must apply to every cluster
observed by EUVE, i.e. {\it all clusters must appear in the EUV like
A2199}.  That this is not the case is immediately revealed by
a comparison of the brightness profiles of A2199 and A1795.  As
Figure 7 clearly shows, they are not consistent with each other.

The evidence presented are sufficient to secure a firm conclusion.
The radial profiles of blank sky and pairs of
cluster fields provided stringent tests of
the integrity of the background subtraction
procedure used by the original authors who announced  the CSE discovery.
The verdict is favorable:
while it is always possible to interpret the data in a less straightforward
way,  any attempt to attribute genuine cluster signals to illusions created
by substantial deviations from flatness of the underlying background profile
must invoke several very artificial arguments.  In fact, the only premise
upon which one can sensibly explain all the data is that
for the clusters in question the underlying
background were reasonably flat, and the correct cluster EUV profiles are
the published ones.

\newpage

{\bf References}
\vspace{5mm}

\noindent
Bowyer, S., Lampton, M., Lieu, R. 1996, {\it Science}, {\bf 274}, 1338--
1340. \\ 
\noindent
~Bowyer, S., Lieu, R., Mittaz, J.P.D. 1998,
{\it The Hot Universe: Proc. 188th \\
\indent IAU Symp., Dordrecht-Kluwer}, 52. \\
\noindent
~Durret, F., Dos Santos, S. and Lieu, R. 1999, ApJ, in preparation .\\
\noindent
~Ensslin, T.A., Biermann, P.L. 1998, {\it Astron. Astrophys.}, {\bf 330},
90--98. \\
\noindent
\noindent
~Fabian, A.C. 1996, {\it Science}, {\bf 271}, 1244--1245. \\
~Fusco-Femiano, R., Dal Fiume, D., Feretti, L., Giovannini, G., Matt, G.,\\
\indent Molendi, S. 1998, {\it Proc. of the 32nd COSPAR Scientific Assembly, \\
\indent Nagoya, Japan (astro-ph 9808012)}. \\
~Kaastra, J.S. 1998, {\it Proc. of the 32nd COSPAR Scientific Assembly, \\
\indent Nagoya, Japan }. \\
~Kaastra,J.S. ,Lieu, R., Mittaz, J.P.D., Bleeker, J.A.M., Mewe, R., \\
\indent Colafrancesco, S. 1999,  { \it Ap.J.L.} in press (astro-ph/9905209). \\
\noindent 
~Lieu, R., Mittaz, J.P.D., Bowyer, S., Lockman, F.J.,
Hwang, C. -Y., Schmitt, J.H.M.M., \\
\indent \it Astrophys. J.\rm, {\bf 458}, L5--7 (1996a) \\
\noindent
~Lieu, R., Mittaz, J.P.D., Bowyer, S., Breen, J.O.,
Lockman, F.J., \\ 
\indent Murphy, E.M. \& Hwang, C. -Y., {\it Science}, {\bf 274}, 
1335--1338 (1996b). \\
\noindent
~Lieu, R., Bonamente, M. and Mittaz, J.P.D. 1999, {\it ApJL} in press.\\
\noindent \noindent
~Mittaz, J.P.D., Lieu, R., Lockman, F.J. 1998, {\it Astrophys. J.}, {\bf 
498},
L17--20. \\
~Rephaeli, Y., Gruber, D. and Blanco, P. 1999 , {\it ApJL.}, {\bf 511},  L21--24. \\
\noindent
~Sarazin, C.L., Lieu, R. 1998, {\it Astrophys. J.}, {\bf 494}, L177--180. \\
\noindent
\end{document}